# Towards knowledge sharing in disaster management: An agent oriented knowledge analysis framework

**Dedi Iskandar Inan**
Papua University, Indonesia
School of Computing and Information Technology
University of Wollongong, Australia
dii740@uowmail.edu.au

**Ghassan Beydoun**
School of Computing and Information Technology
University of Wollongong, Australia
baydoun@uow.edu.au

**Simon Opper**
Emergency Risk Management Branch
SES New South Wales, Australia
simon.opper@ses.nsw.gov.au

## Abstract

Disaster Management (DM) is a complex set of interrelated activities. The activities are often knowledge intensive and time sensitive. Sharing the required knowledge timely is critical for DM. In developed countries, for recurring disasters (e.g. floods), there are dedicated document repositories of Disaster Management Plans (DMP) that can be accessed as needs arise. However, accessing the appropriate plan in a timely manner and sharing activities between plans often requires domain knowledge and intimate knowledge of the plans in the first place. In this paper, we introduce an agent-based knowledge analysis method to convert DMPs into a collection of knowledge units that can be stored into a unified repository. The repository of DM actions then enables the mixing and matching knowledge between different plans. The repository is structured as a layered abstraction according to Meta Object Facility (MOF). We use the flood management plans used by SES (State Emergency Service), an authoritative DM agency in NSW (New State Wales) State of Australia to illustrate and give a preliminary validation of the approach. It is illustrated using DMPs along the flood prone Murrumbidgee River in central NSW.

**Keywords:** Agent Oriented Analysis, Metamodelling, Disaster Management, Knowledge analysis

## 1 Introduction

A disaster event is unpredictable and uncertain (Ramete et al., 2012; Scerri et al., 2012; Wex et al., 2012). Managing activities involved can be very challenging and complex (Helbing, 2013). For a given DM activity, there are often numerous stakeholders: agencies, organizations, and individuals with various roles and different backgrounds, resources and goals (Wang & Hsiao, 2014). Timeliness of action is also critical (Janssen et al., 2010). As the nature of DM is a complex system, the knowledge exchange among entities involved in the DM activities is not only very critical but also it has to put in place in time (Heard et al., 2014). Particularly, in a DM activity those entities bring their own structures and background and they need to be communicated and negotiated among them with respect to a resilient agenda of DM activities. An authoritative DM agency as the leading agency to combat the disaster will take a lead to organise and elicit the knowledge which ultimately structures it in to a shareable and reusable format to entities in any DM activity. However, accessing the knowledge which is specified in a semi-structured natural language format is very challenging. The written knowledge tends to be structured in a business specification format which in fact, is subjective to the entities and need a deductive process to get insight from. Moreover, the knowledge could be scattered in the document (DMP) that will be frustrating to the readers particularly in an imminent disaster.

In this paper, we view the challenge of DM as one of harnessing and sharing knowledge between agents to timely and effectively reduce the impact of a disaster. The first step towards this is codifying DM knowledge to facilitate its reuse and sharing. However, analysing the knowledge in a complex domain, such as DM, is not only difficult but also time consuming (Brown et al., 2014). In Australia, the PPRR (Preparedness, Prevention, Response, Recovery) model is typically used to organise DM knowledge (Rogers, 2011). Various DM activities and knowledge units required throughout the DM processes are organized according to the sequence of four phases: Prevention (P), Preparedness (P), Response (R) and Recovery (R). With all its prominence, PPRR has been criticized for not conceptualizing the process of disaster management holistically, rather it does it sequentially (Rogers, 2011). This has been explained as a relic feature of PPRR predating the modern view of aiming to have risk management permeate all DM activities (Crondstedt, 2002). A linear and a sequential description of events are inherently limited. They do not allow participants to engage beyond the tip of the event timeline. In software processes this sequential modelling has been abandoned many years ago to mitigate against the risk of introducing software errors (Lopez-Lorca et al., 2015). It is well accepted that software practitioners typically engage in iterative thinking and problem solving, moving up and down multiple abstraction layers. Applying this same paradigm and insights to representing disaster management processes, we adopt a multi layered metamodelling approach which follows MOF (Meta Object Facility). As a first knowledge analysis step to enable this, the paper proposes an approach based on Agent-Oriented Analysis (AOA) to codify DM knowledge appropriately.

Disaster Management Plans (DMP) do not articulate a single goal. Entities involved in a DM activity need to not only react or adapt to the environment but to also exhibit their local goal formulation (Doyle et al., 2014). Critical environment characteristics can't be controlled and predicted but awareness of them is critical to facilitate cooperation. Entities/organizations/individuals involved have their own goals, resources and structures. At the same time, the need to communicate and negotiate to pursue common goals is paramount. This creates an imperative for timely sharing and reusing of knowledge. This paper advocates the use of a knowledge repository based on a common MOF modelling framework, Disaster Management Metamodel (DMM) (Othman et al., 2014). Specifically, the paper addresses the challenge on how to convert existing DM knowledge into DMM based constructs. This enables layering of abstractions and abandoning a timeline sequence, in favour of free flow access of any point. The challenge addressed in this paper is how to convert end user models to concepts and notation from DMM. The paper deploys AOA towards this. Agent oriented models lend themselves to represent organizational know-how and DM processes. They emphasize the constructs of roles, agents and organizations to represent systems behaviours. Much know-how and processes in DMPs across Australia are actually expressed in such terms. With appropriate supporting tools, this knowledge can be deposited and shared using a DMM-based system.

The rest of this paper is organized as follows: The next section reviews the background and related work. The third section presents our intermediate framework of AOA to convert extant DMPs domain knowledge to DMM constructs. Fourth section illustrates the approach using actual NSW State Emergency Services DMPs. Finally the paper concludes with a discussion of future work.

# 2 Related work

Metamodelling creates a collection of classes to describe domain concepts to represent domain entities, actions or states (Othman & Beydoun, 2013). The collection of concept is a metamodel which contains the specification of modelling environment and defines the syntax and the semantics of the domain (Syriani et al., 2013). Classes and relations in a metamodel represents complete constructs and rules of knowledge in real world (activities, interactions, conditions, actors, roles, triggers and so on) in the conceptual level from a particular domain. The development process complies to the rigorous methodology of Model Driven Engineering (MDE) (Whittle et al., 2014). Likewise, the DMM developed by (Othman & Beydoun, 2013) is a metamodel representing all concepts and relations in DM domain as it is developed with respect to a 98-DM model from government, private and academic DM models. Similar constructs and rules from those models are reconciled and validated to produce a generic one called DMM. As the objective of a metamodel is completeness (Beydoun et al., 2009b) thus the developed DMM also represent all common concepts exist in the DM domain. For instance a concept: *PublicEducation* in Preparedness phase of the DMM (see Figure 1), the terminology of the concept is: "*A process of making the public aware of its risks and preparing citizens for hazards in advance of a disaster and as a long-term strategic effort*" (Othman, et al., 2014, p. 28). This implies that all the real knowledge in a DM activity related to public education in Preparedness phase will be dragged in to the concept in the conceptual level, and so on. However, since DM is a complex domain, analysing the knowledge from the real world to be able to see and extract the specific characteristics of the domain requires a suitable methodology which can cope which all those complex characteristics. In addition, as regards MOF framework, once the real word characteristics are analysed and in place, the transformation process should follow the MOF framework in MDE methodology to ensure the knowledge is tailored to its conceptual level correctly. It implies the real work activities in a DM can only be mapped to their conceptual levels in the DMM through modelling processes in between and vice versa. For instance, real word knowledge in Preparedness phase regarding *"…responsibilities to ensure the residents in the council area are aware of the flood threat in their vicinity and how to protect themselves from it"* (SES NSW Australia, 2006, p. 14). This activity is intertwined with so many other activities, for instance, who are involved in, when they should be performed, what resources are required, what are the pre- and post- conditions of the activity and so on, to be able to be mapped to their appropriate concepts and relations in the DMM. Therefore, an intermediate modelling activity is required to enable the knowledge transform from the real activities to its appropriate concepts and relations in the conceptual level.

Replacing PPRR with a rigorous metamodel generalizes practices and enables partitioning of DM problems into sub-problems easier to tackle. It can also provide an easily accessible layered representation of knowledge. This can subsequently enable stakeholders to engage at all levels of abstractions as required (events, policies and organisational structures). Various works recently use metamodeling to represent diffused DM knowledge e.g. (Chen et al., 2015; Lauras et al., 2015; Ramete, et al., 2012). However, most focus on specific DM phases (e.g. either Prevention/Mitigation, Preparedness Response or Recovery). Unlike this paper, none yet provides any support for converting the DM knowledge into the metamodel constructs themselves. The paper deploys a generic broad DM metamodel (DMM) that is disaster and phase independent. The work uses an existing metamodel, DMM (Othman & Beydoun, 2010), to illustrate the proposed agent oriented approach. DMM is partially shown in Figure 1 (Only 1 phase of a DMM is shown here to accommodate the space limit).

As practitioners are engaged in responding to a disaster, their actions are event driven however their reflections and motivation may be policy driven or even constrained within the structure of their organization. In other words, knowledge generated during the events pertains not only to the event, but also to the policy development and perhaps reflections on scope for restructuring. Enabling the representation of this abstract knowledge is paramount to enable continuous process improvement. PPRR used in Australia is limited by capturing only one perspective at a time and furthermore it assumes sequencing of the activities of Prevention, Preparedness, Response, and Recovery (PPRR) (Seidita et al., 2010). DMM has been developed as a representative set of concepts and relations in a DM which can be used to store DM knowledge from real DM activities.

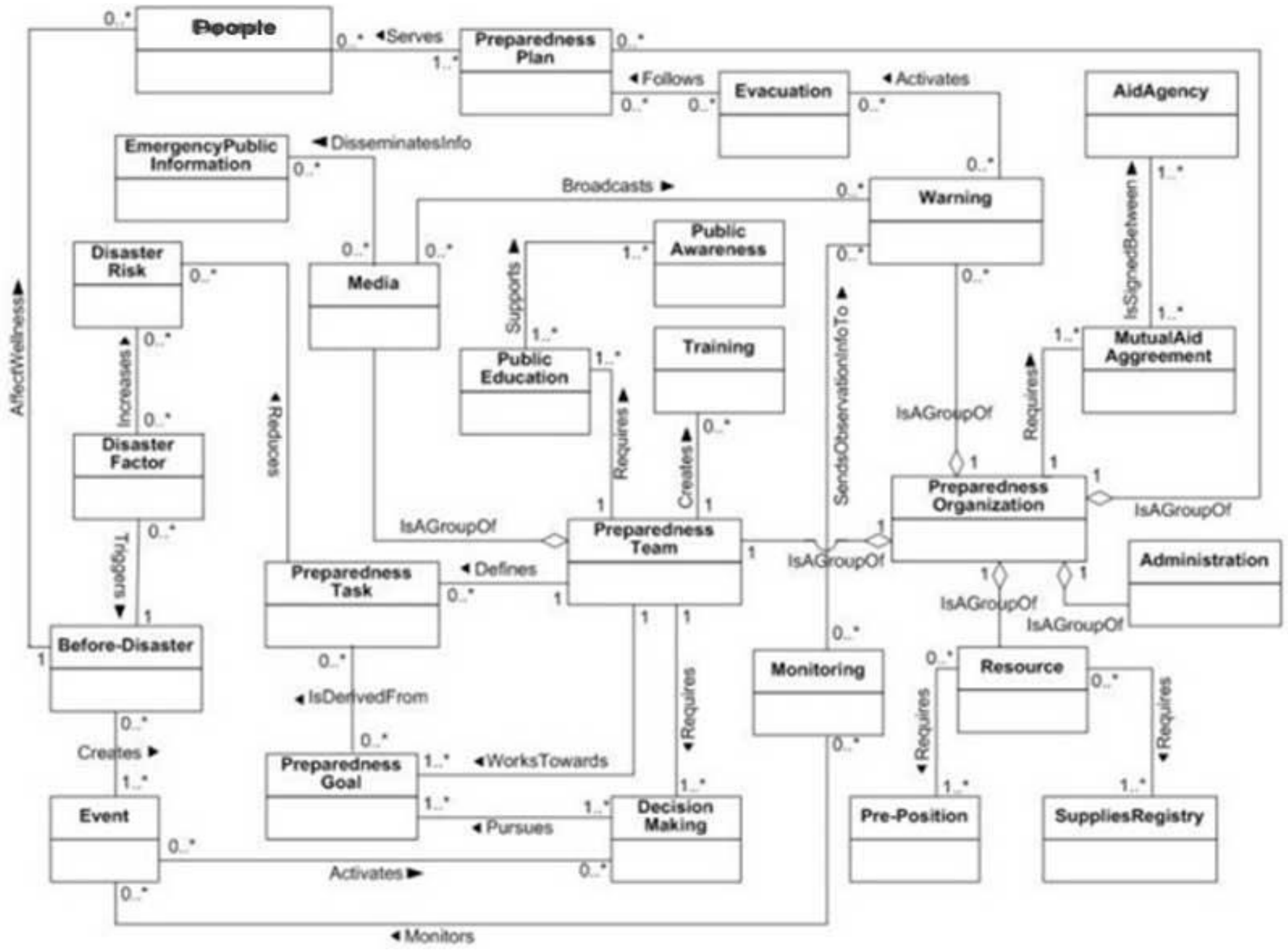

*Figure 1. Preparedness Phase of DMM*

DM modelling aims to capture the complex characteristic of DM and present it in a way common people can understand easily of the domain (Sackmann et al., 2013). DM has these characteristics; a) Situatedness in an environment (Cavallo & Ireland, 2014). As disasters are dynamic, unpredictable and uncertain, an environment changes rapidly which lead to the second characteristic. b) Time sensitivity (Janssen, et al., 2010), in a disaster, every activity has to deal with deadlines otherwise the consequences might lead fatalities to casualties; c) Non-deterministic (Wex et al., 2014). Disasters often throw unexpected eventualities. This factor means the unpredictability is very high. d) Presence of autonomous entities (Ernstsen & Villanger, 2014). This means that in a DM activity, individuals/agencies/organizations are coming from different backgrounds, knowledge, abilities, structure, mandate, no common perception and so on. Agent-Oriented (AO) modelling approach enables analysis of complex systems, in particular socio-technical systems (Sterling & Taveter, 2009). AOA is easy to understand for humans, as it uses constructs from familiar organisational setting (e.g. roles, activities, interaction and etc.) (Miller et al., 2014). It is at a high-level of abstraction that enables analysts to apply concepts that they are familiar with from their daily deductive processes (Winikoff & Padgham, 2013). There are clear similarities between AOA and the context of DM, most strikingly: agents are driven by local goals and need to interact towards a system goal; agents have specified roles and need to interact accordingly; agents are situated and need respond in real time in many instances (Lopez-Lorca et al., 2011b). Not surprisingly, there have various attempts recently to use AOA to support DM e.g. (Aldewereld et al., 2011; García-Magariño & Gutiérrez, 2013; Padgham et al., 2014; Scerri, et al., 2012). However, much of these works focus on developing simulations of disaster events to gauge the effectiveness of existing practices. This paper applies AOA templates to convert disaster management to an intermediate form which can then be converted to DMM based constructs.

The paper introduces a knowledge analysis framework based on AOA to facilitate modelling and sharing of DM knowledge. The analysis framework includes activities to annotate the concepts to facilitate mapping to AOA constructs. This in turn enables the conversion of DMPs to the shareable form that enables DM stakeholders in cooperative decision making processes. The analysis framework in essence bridges the semantic gap between the unifying DMM and existing DMPs. The framework is then illustrated using actual flood management disaster plans of State Emergency Service (SES) of New South Wales in Australia.

# 3 An agent-based knowledge analysis framework

In a typical Disaster Management Plan (DMP), a set of goals state the intent of Disaster Management (DM) for Preventing and Preparing for, Responding to and Recovering (PPRR) from disasters. A particular DMP will guide the participants in that plan to a set of problem-solving tasks required to be pursued by contributing organisations, groups and particularly by Incident Management Teams. These participants will be located in specific areas of authority and have hierarchical levels of control and command such as the NSW State Emergency Service (NSW SES) which is the legislatively appointed combat (lead) agency to plan for and control flood, storm and tsunami disaster management operations. This is implemented through NSW SES Local, Region al and State organisational levels during day to pre-disaster planning and also by specific incident controllers in Incident Management Teams during response. However even within this construct hierarchy and control complexities exist, for example while the NSW SES is the combat agency for flood disaster management, a NSW Police Commander will control specific tasks for which NSW Police is the controlling or lead agency. An enacted emergency plan requires all involved to be well conversant with potential tasks required in the PPRR cycle. This is reflected in the Total [Flood] Warning System (1) which includes but not limited to public information and warning, staff and volunteer mobilisation, evacuation, rescue, intelligence (2), situation awareness and planning strategies to protect elements ( infrastructure or community) at risk. Knowledge of the relation between various tasks and how the specific area of control overlaps with adjacent organisations but particularly between Incident Management Teams at Local, Regional and State levels is an essential part of the success in implementing the DMP. Accessing this knowledge leads to a cascade of further context awareness. It typically leads to further identification of other related-knowledge along with those tasks which might be performed sequentially, parallel or interleave. And in terms of performing those tasks, an agent (a person, a group of people or an agency) may play various roles and interact with numerous other agents. Furthermore, agents typically have different scope and belong to different layers in various administrational or command and control hierarchies. Withstanding this, the agents still need to be able to communicate to each other to pursue a particular goal(s). As they collaborate, agents are often required to maintain their own situation awareness and need to react to changes in their environment as events unfold. In the midst of all of this, agents need to be knowledgeable of not only their goals but also of their resources and supporting systems.

The breadth and complexity of this knowledge presents a number of significant challenges for disaster managers and participating organisations as well as the community. The NSW SES prepares and maintains some 123 individual Local Flood Plans across NSW Local Government Areas and this involves extensive processing of flood risk data and consultation with all organisations and participants involved to develop strategies in the plan. Other hazard managers such as bushfire maintain similarly large numbers of Local and Regional level disaster plans. While there are multiple issues which can be benefited by the work outlined in this paper, for example improving the inefficient maintenance of such a large connected but disparate knowledge representation currently maintained as individual Microsoft WORD documents, the critical outcome discussed in the paper is the importance of shared understanding and ease of access to disaster management knowledge, roles and actions. For example how is a participating organisation or officer, or an individual in the community best enabled to explore and understand their role and actions in the context of a large and complex disaster management plan? A resilient community is one which has awareness of its risk and of strategies to deal with it before disasters strike, then enacts this during disasters when there will be little time to try and develop this understanding for the first time from large and complex documents. This challenge is the core theme of this research.

Analysis and sharing of the knowledge above requires a systematic approach to structure the knowledge and communicate it easily (Bera et al., 2011). The analysis requires answering complex questions such as: how a goal can be identified and evaluated; how agents negotiate their priorities as they collaborate in common goal(s); what specific activities agents perform as they pursue their goal(s); what resources are needed for given goals or agents; what time and resource constraints should be imposed on particular agents; and so on. The proposed framework of knowledge analysis of a DM domain within a DMP transforms the knowledge involved into a representative repository to enable reuse and sharing.

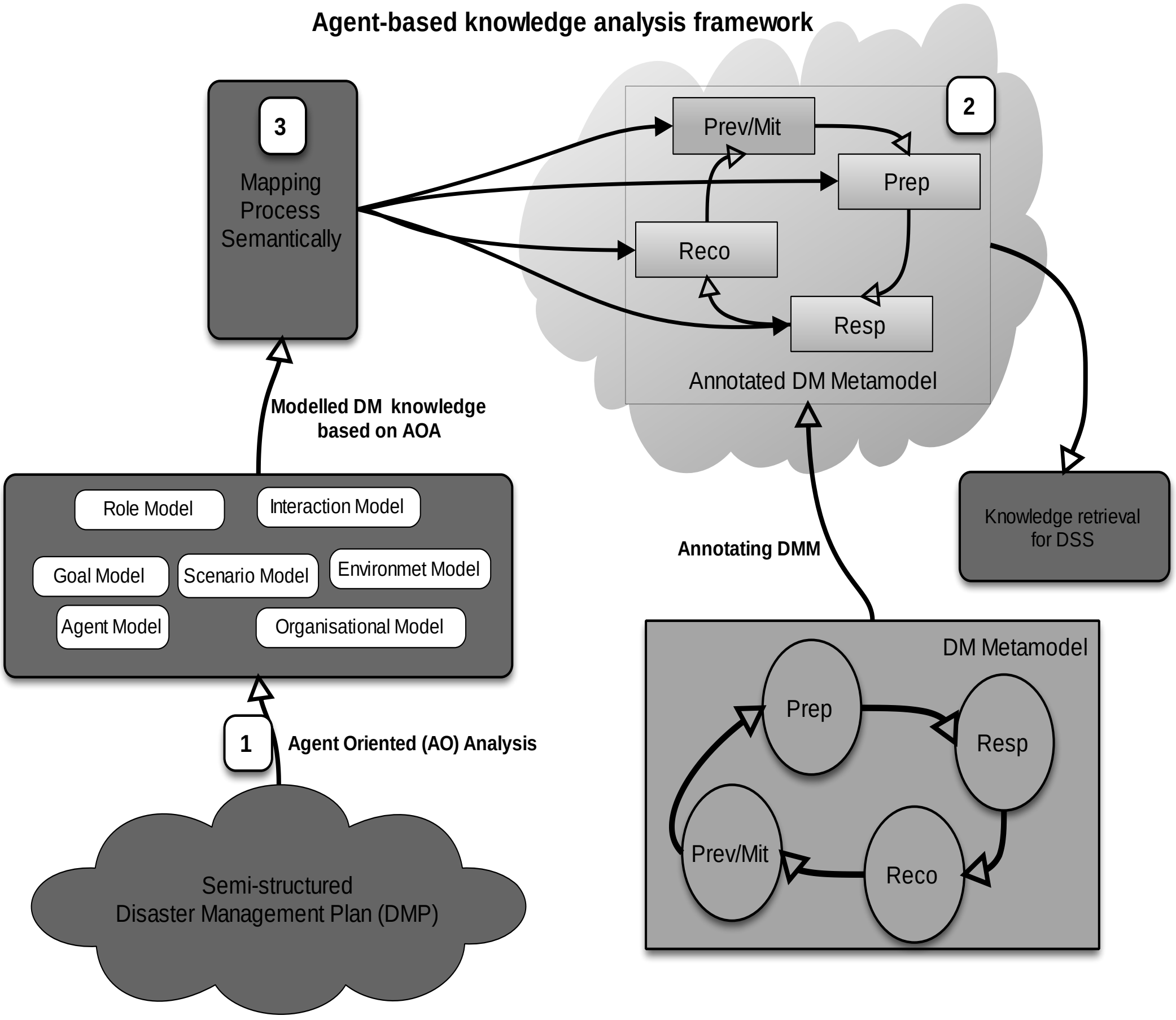

*Figure 2. Overview of the conversion process of DMPs into DMM based knowledge repository*

Our analysis framework is shown in Figure 2. It consists of three steps: Step 1 is the agent-based analysis which applies these models (templates) to the DMP. Step 2 formulates a mapping between the agent model components and DMM concepts. Step 3, transfers the DM knowledge from the agent models to a repository of DM knowledge structures using the generic DM metamodel (DMM). The mapping of Step 2, underpinned by an annotation process of DMM concepts, facilitates the transformation process of Step 3. At the end, a DM knowledge repository is available to support DM decisions. The remainder of this section details each of steps in our knowledge analysis framework.

## Step 1: Agent-Oriented Analysis (AOA) of DMPs

Agent-oriented models can represent organisational processes and activities as described in a typical DMP. This step transforms the semi structured DMP specifications into a set of agent models to be later converted into DMM constructs. Concepts used in AOA processes are organized in models that are accessible to many stakeholders in DM. We identify the following set of seven AO templates to capture the knowledge from DM: goal model, role model, agent model, organisational model, interaction model, environment model and scenario model. The details of these models follow the work in (Lopez-Lorca et al., 2011a) and are as follows:

1. A *goal model* is introduced to capture the reactiveness and proactiveness knowledge of the agents involved in the DM. In this model, roles that need to be played to achieve the goal(s) are identified. The sub-goals as subsets of the goals are also identified.
2. A *role model* is used to represent all responsibilities knowledge that needs to be played by an agent and all the constraints of those responsibilities of a role.
3. An *organisational model* is used to represent the relationships between roles and highlight how to take in to account their relationships in a DM process. The model defines the communication channels between agents which may belong to different organisations or levels of command in a widely dispersed disaster.
4. An *interaction model* is used to elaborate the specification of the communications between agents that play particular roles to pursue a goal. In other words, this model defines in which goal agents need to interact.
5. An *environment model* specifies the environmental constraints on activities and resources of agents. This model elaborates the resources, the activities and the roles that require them.
6. An *agent model* elaborates the type of agents involved in DM activities, along with further specifications of their activities and their goals. Triggers are also identified to represent event(s) that spur agents into action. This is the manifestation of an agent's situatedness in an environment.
7. A *Scenario model* binds all knowledge fragments as activities that need to be undertaken in pursuing a particular goal with specific triggers and agent types. The activities are preceded by a pre-condition and followed a post-condition, as a desired state of the goal that pursued in the activities. Conditions of those activities are specified as either parallel, interleaved or sequential.

## Step 2: Annotating DMM concepts with AO concepts

This step prepares the structure of the repository to receive the agent models from Step 1 (see Figure 2). This step readies the repository to receive the agent based models into DMM constructs in Step 3. The repository is structured using the metamodel, DMM, which was synthesized using 89 existing DM models and covers all common concept across all those DM models (Othman, et al., 2014). This step provides the basis of a mapping between the elements of the models from Step 1 and the DMM constructs (shown in Figure 2). DMM abstractions layers are consistent with MOF. To ensure the mapping is consistent along the abstraction layers defined by MOF, we also use a corresponding agent metamodel, FAML (Beydoun et al., 2009a), which describes agent models at MOF abstractions as well. The FAML metamodel is used as the basis annotating process of DMM. Appropriate Agent-Oriented Software Engineering (AOSE) metamodel concepts are mapped to each agent model used in Step 1. This mapping then provides a set of terms that are used to annotate DMM appropriately. This annotation process is a one-off process for all concepts in the DMM with their appropriate annotation concepts from AOSE metamodel construct. An example of this process is given in Table 1.

Table 1 shows all the agent-based models in Step 1 in their metamodelling level but organisational model and interaction model. In AOA from Step 1, organisational model is aimed to represent the hierarchy level of all roles played agent(s) to be able to communicate each other accordingly. This means by having this knowledge then each role played by an agent knows how to approach and communicate to other roles played by other agents in the different administration level of hierarchy, whether they are in the same level, lower level or higher level. In the DM context, the role played by an agent might also belong to a different scope, for instance, a national or international aid agency. This sort of knowledge is captured in organisational model. To represent the hierarchy level in the higher layer of the modelling process, an AOSE metamodel, these relationships are added: *isPeer* relationship presents roles played by agents in the in the same hierarchy level, Controls and *isControlledBy* represent authority relationships between agent roles. The interaction model represents what goal(s) agents are cooperating towards and their relevant interactions. In the higher layer of modelling process with respect to MOF framework, an interaction model is represented by a relationship class *rolePursueGoal* defined as a role(s) played by an agent to pursue a particular goal(s). Therefore for the two agent-based models, organisational model and interaction model, are in the relationship of the exiting constructs.

| DMM Concept in Preparedness phase | AOSE metamodel Constructs | Description |
|---|---|---|
| Preparedness Goal | <<Goal>> | Represents a certain condition need to be achieved by the system |
| PreparednessTeam | <<Role>> | Represents a set of capabilities to perform by agent to achieve the goal(s) |
| PreparednessTeam | <<Agent>> | Represents an entity that having certain properties and can play one or more role |
| Training<br>PublicEducation | <<Activity>> | Describes a set of activities to be performed to achieve the goal(s) |
| Before-disaster | <<Event>> | Defines a situation change that influence a significant change of an agent to respond the situation |
| Media<br>MutualAidAgreement | <<EnvironmentEntity>> | Represents any resources required to perform the tasks |

*Table 1. Examples of AOSE metamodel Concepts mapped to DMM concepts*

A knowledge modeller is required to annotate each DMM concept with the appropriate AOSE concept as described in Table 1. The Training concept for example, is defined as follows: "*An instruction that imparts and/or maintains the skills (and abilities such as strength and endurance) necessary for an individual, a community or an organization to perform their assigned disaster action responsibilities*". This is a set of activities to be undertaken to maintain the skills of DM stakeholders. This consists of a set of activities hence the corresponding concept from the AOSE metamodel is <<*Activity*>>: "*Describes a set of activities to be performed to achieve the goal(s)*". Therefore the modeller annotates Training concept in DMM with the <<Activity>>. Another example is a *PreparednessTeam* defined as follows: "*A group of all agencies with a role in incident management that provide interagency coordination for domestic incident management activities in a non-emergency context to ensure the proper level of planning, training, equipping and other preparedness requirements within a jurisdiction or area*". This concept describes a set of roles played by an agent(s) to pursue a goal(s) in a DM activity. As a role is representing a set of capabilities is played by an agent, the AOSE appropriate concepts in the metamodel for *PreparednessTeam* are <<*role*>>: "Represents a set of capabilities to perform by agent to achieve the goal(s)" and <<agent>>: "*Represents an entity that having certain properties and can play one or more role*". Therefore, a knowledge modeller annotates the *PreparednessTeam* to the <<*agent*>> concept and <<*role*>>.

DMM has 92 concepts: 21 concepts in Prevention, 25 concepts in Preparedness, 25 concepts in Response and other 21 concepts in Recovery. More than one concept in DMM can get mapped to the same AOSE metamodel concept. For example, many concepts in DMM are about activities and resources/supporting systems. Hence, agent concepts of <<Activity>> and <<*EnvironmentEntity*>> are repeatedly mapped to many concepts in DMM. From Table 1, *Training* and *PublicEducation* concepts are annotated as <<*Activity*>>, MutualAidAgreement and Media are annotated as <<*EnvironmentEntity*>>. The annotation of DMM with AO concepts is a one-off process. However, the knowledge modeller can revisit the mapping process if required.

## Step 3: Transferring DMPs AO models into DMM-based repository

In this step, agent-based knowledge models acquired from DMPs are transferred into DMM representation using the mapping of Step 2.

This process is the foundation of our intermediate framework of knowledge analysis as it converts knowledge from a lower layer to its higher layer. It transforms user knowledge to its metamodel (M0-M2, with respect to MOF framework). By adopting MOF in software engineering, tangled knowledge of DM can be pinpointed to which abstraction layer it belongs. The activities in this step are undertaken semi automatically. A DM practitioner is involved in transferring the models to their appropriate DMM constructs. From step 2, the AOSE concepts (as depicted in Table 1) map to 92 DMM concepts across all phases in a DM. An AOSE concept maps to multiple DMM concepts. The DM practitioner selects a subset of the possible DMM constructs. They judge semantically which concept in DMM are appropriate to capture the information in the agent-based knowledge models. Table 2 depicts mapping processes of agent-based knowledge models from AOA stage to their appropriate annotated DMM concepts. With respect to the MOF hierarchy, not all agent based models are represented equally. Some models generate more constructs at M0 level than others. Other models generate more constructs a M1 level. For instance, Scenario and agent models generate more constructs at M0 whilst Role and Goal models generate more constructs at M1 level.

| Agent-based Knowledge Model | Description | Annotated DMM Concept |
|---|---|---|
| Scenario model | Knowledge related to training activities | <<Activity>>:Training |
| Scenario model | Knowledge related to public education activities | <<Activity>>: PublicEducation |
| Agent model<br>Role model | Knowledge related to a role played by an agent | <<Agent>>,<<Role>>: PreparednessTeam |
| Environment model | Knowledge related to medias as supporting systems | <<EnvironmentEntity>>: Media |
| Environment model | Knowledge related to mutual aid agreement as supporting systems | <<EnvironmentEntity>>: MutualAidAgreement |

*Table 2. Mapping process of agent-based knowledge models to the appropriate annotated concepts*

To validate the overall knowledge analysis framework, the process is validated in converting NSW State Emergency Services DMPs to DMM constructs. The DMM-conversion focusses on DMPs from Wagga Wagga, Murrumbidgee Central NSW Australia. A DM practitioner from SES (the third author) is involved in this process. This case study described in the next section.

# 4 CASE STUDY: Agent-based conversion of WAGGA WAGGA FLOOD DISASTER PLAN

As early as the first European settlements on the Hawkesbury River in Sydney, development pressure in flood risk areas has exposed people and communities to flooding and resulted in deaths and high damage costs. Flood deaths rank second behind heatwaves for natural hazard fatalities in Australia (Gissing et al., 2010) and the cost of disasters generally is increasing by tens of millions dollars per year. Climate change modelling suggests that while there are likely to be little changes in average rainfall across the state by 2030, and there will be large seasonal differences. The frequency of coastal flooding may increase as a consequence of sea level rise and potential increased frequency of storm surge events, particularly as the events coincide. Risks to population and infrastructure are likely to increase as a consequence of sea level rise and the increased severity and frequency of storms and coastal flooding (NSW Government, 2013).

Amidst this backdrop of rarer but more severe weather is the reality that overall exposure of people and infrastructure is increasing from ongoing development and population increase. Disaster risk is going up, not down and risk management is about minimisation, not overall net reduction. Also that even while future disasters may be generally rarer leading to decreased awareness, already communities have little contemporary knowledge of disasters, especially large disasters which have occurred in the past but just outside the current life of those alive today. Thus a central concern for communities and all levels of Government is how to assure that we have learnt from past experience and planned for the future to create more flood resilient communities.

The regional town of Wagga Wagga (WW) and surrounding rural area, in the City of Wagga Wagga Local Government Area, NSW, is situated on the Murrumbidgee River floodplain, the second longest river in Australia. The history of Flooding in Wagga Wagga is a good example of the sporadic frequency of flooding in inland Australia being the driest inhabited continent on Earth. The sporadic nature presents major challenges for maintaining community and Government awareness and knowledge of flooding and of ongoing flood resilience with large periods of drought between major floods.

Flood disaster management in New South Wales is coordinated through a set of documented emergency/disaster plans and arrangements at the Local, Regional and State levels. The Wagga Wagga

Local Flood Plan (LFP) is a flood hazard specific sub plan supporting a Regional Disaster Plan (DISPLAN). Other sub plans focusing on Health, Agriculture and Energy and Utilities etc., also support the Regional DISPLAN and are enacted during disasters such as floods. The Regional DISPLAN is in turn a sub plan supporting the State Emergency Management Plan (EMPLAN).

The Wagga Wagga LFP is maintained to prepare for, manage the response to and support recovery from flood disasters. It is maintained by the NSW State Emergency Service in conjunction with the Wagga Wagga City Local Government and their representative Local Emergency Management Committee comprised of local stakeholders. The plan can be downloaded freely from SES (State Emergency Service) website: http://www.floodsafe.com.au/. In the context of this paper, the LFP is considered as a semi-structured document, as the knowledge in it has been populated and written in a particular style and structured by practitioners involved in the DM for floods. It covers knowledge in three phases: Preparedness, Response and Recovery. However, for this paper, the modelling process is applied only in the Preparedness and Response phases as each of them has represented pre- and post-disaster phases of DM. The process of converting DM knowledge into its repository, DMM is as follows, the process consists of three main activities: (1) Analyse DM knowledge based on AO analysis; (2) Annotating concepts of DM metamodel; and (3) Semantic mapping process of agent-based knowledge to its appropriate concept in metamodel. These activities are called the intermediate framework as it will describe the step by step process to intermediate the storing process from agent-based knowledge models to the metamodel.

## Step 1: Knowledge Modelling based on AO analysis

In the first step, analysis processes are applied to the semi-structured knowledge of DMP using the seven agent-oriented templates described in the previous section. In this step, knowledge of Preparedness and Response phases from Wagga Wagga DMP is reformulated using the AO. This is a manual labour intensive step. This process is performed iteratively to ensure that all knowledge in the DMP has been appropriately captured.

The modelling process begins with the producing using the goal models. These represent the purposes that are intended to be achieved in all activities in a DMP. This model also identifies role(s) played by agent(s) involved in the activity to pursue the goal. In this case study, the goal models are applied to model the knowledge from Preparedness and Response phases of the DMP. Once these models are considered mature the next model template is used. The next template can be one of these four: the role model, the organisational model, the interaction model or the environment model. The Agent model and scenario model can only be used after the environment entities are identified. This occurs in developing the environment model.

Some iteration is required to make sure that all the knowledge has been analysed and all connections between the seven models have been made and identified. For instance, a knowledge fragment of actions in an agent model needs to refer to corresponding activities in the scenario model, which in turn should have corresponding sub-goals in a goal model. Another example is identifying responsibilities in a role model. Responsibilities should also be identified in a goal model. Thus, iterations over an evolving set of models are required. With each iteration newly identified knowledge may require the modeller to revisit earlier versions of say the goal or role models.

## Step 2: Annotated DMM

In this stage, as described in the previous section, there are 6 AO concepts that are used to tag 50 DMM concepts (25 concepts in each Preparedness and Response phases respectively). Many DMM concepts which will have the same tag (AO concept). Figure 3 and Figure 4 show the annotated DMM concepts in Preparedness and Response phases respectively, as used in the Wagga Wagga case study. A goal model will be mapped with a corresponding goal concept through <<*goal*>> to represent goal to be pursued, a role model will be mapped with a <<*role*>> concept, *environment model* with an <<*environemntEntity*>> concept and so on (see Table 1). To describe hierarchy level among agents involved in a DMP as described in *organisation model* then domain properties of the agent are added as: *isPeer*, representing agents in the same hierarchy level, *Controls* and *IsControlledBy* represent an agent controls another agent or is controlled by others.

Interaction in interaction model between agents to pursue goal(s) is described by adding the relations: *ParticipatesIn*, that describes an agent is participated in a particular activity or in pursuing activity Involves an agent. For instance, an agent A plays a role X and an agent B plays another role Y then the interaction between agents A and B will be shown in their cooperative pursuit of goal P. This is described using the relationship *ParticipatesIn* to achieve goal P; or in another way, goal P Involves agent A and B.

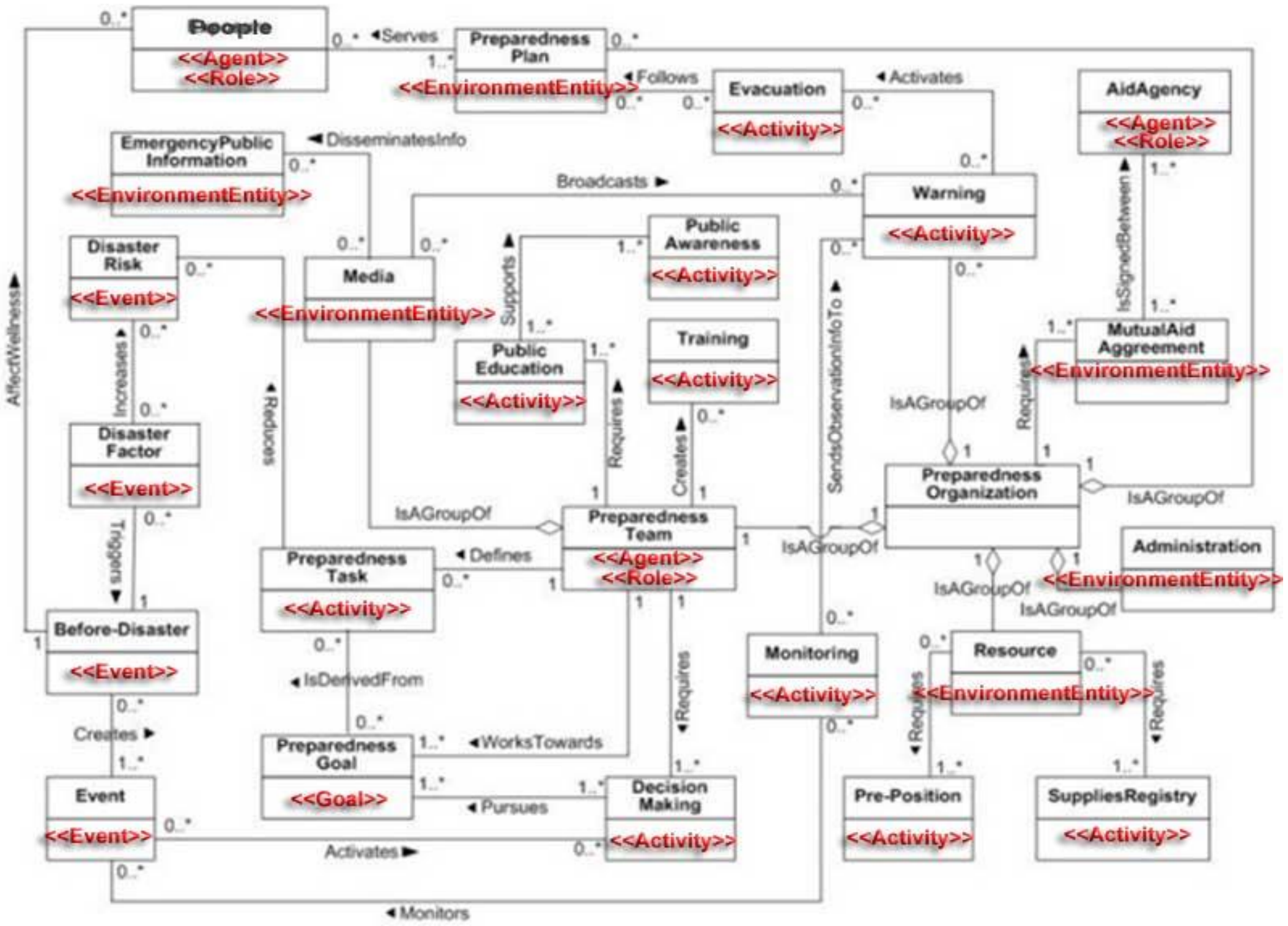

*Figure 3. Preparedness-phase of DMM annotated with AO concepts*

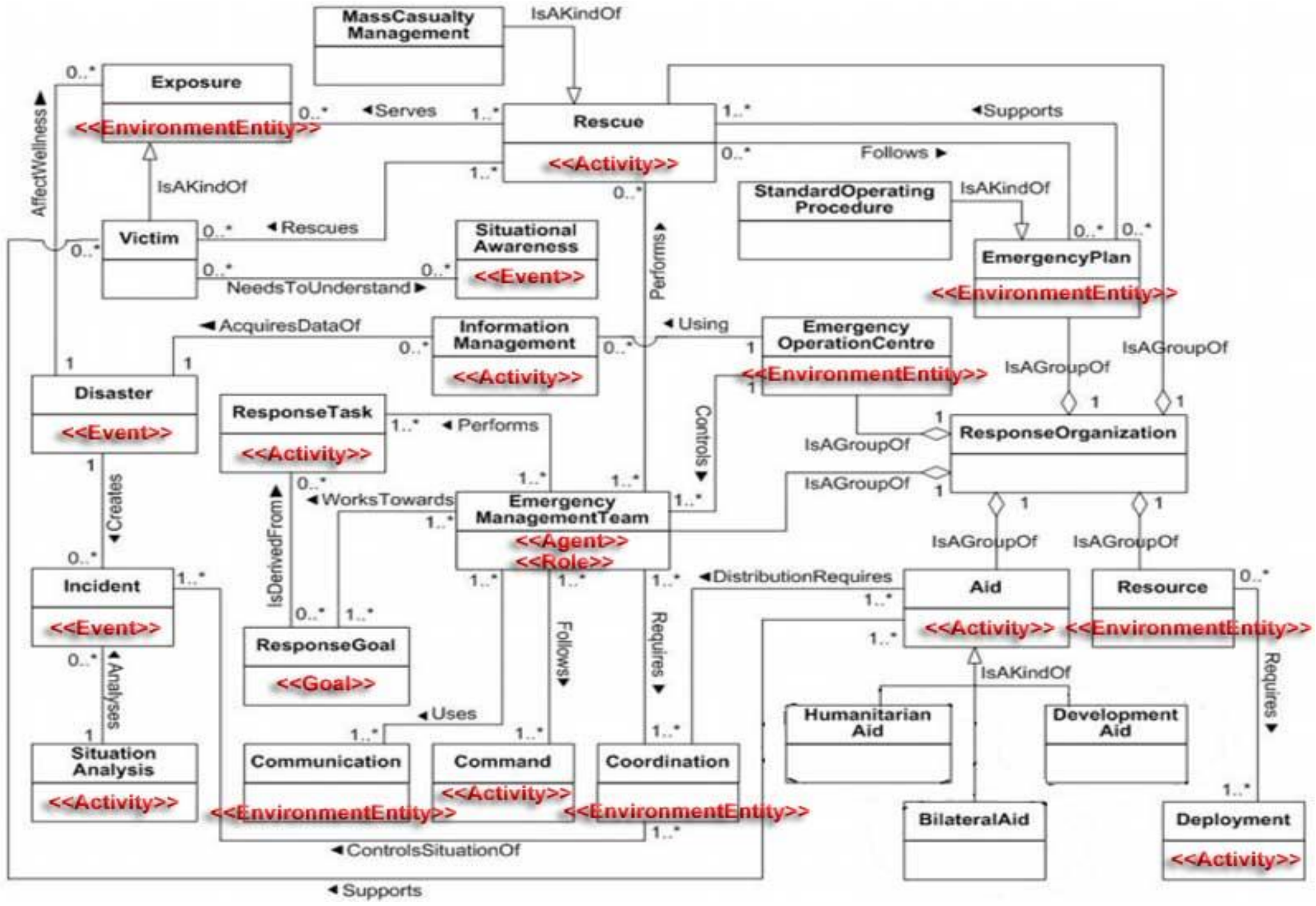

*Figure 4. Response-phase of DMM annotated with AO concepts Response-phase*

## Step 3: Semantic Mapping Process of the DM knowledge

Once agent-based knowledge models and the annotated DMM concepts are in place then mapping the agent models to DMM constructs is a semi-automated process. It still requires a DM practitioner. This process is implemented as graphical web-based user interface that supports access to the DMM knowledge repository. The agent models are made available in XML and they are input to MySQL database. MySQL is a powerful, widely used as well as an open source database and as it harnesses a web-based technology to connect client requests to the server then Apache is a web server technology used in the most web servers around the world. Figure 5 is a web-based interface of mapping process semantically of agent models to their appropriate annotated DMM concept.

Figure 3, it depicts that based on the annotated DMM in Step 2, there is only three concepts related to the roles played by the agent(s) in the DMM, that is *PreparednessTeam*, *AidAgency* and *People*. Therefore the system will automatically limit those candidates of DMM concepts appearing in the list of appropriate annotated DMM concept) (see the Figure 3). This assists the DM practitioner in having broad awareness but concise visibility of the related knowledge, processes and concepts that they must apply within the DMP and the overarching DMM. This is shown Figure 4, where only the list of knowledge about roles played by the agent(s) is mapped to those three candidates DMM concepts (Figure 5). This process is applied in all mapping activities between all agent-based models and the corresponding annotated concept of the DMM metamodel. Thus the DM practitioner, no matter what role they perform will have a much clearer understanding of their responsibilities in the DMP.

# 5 SUMMARY AND FUTURE RESEARCH

In this paper, we present a DM knowledge analysis framework which enables the conversion of Disaster Management Plans (DMP) to a metamodel-based representation in order to facilitate DM knowledge sharing and reuse. The metamodel is 3D knowledge structure that has a 3-layered representation. Each layer corresponds to a more abstract view of DM knowledge, from event, to policy to the metamodel.

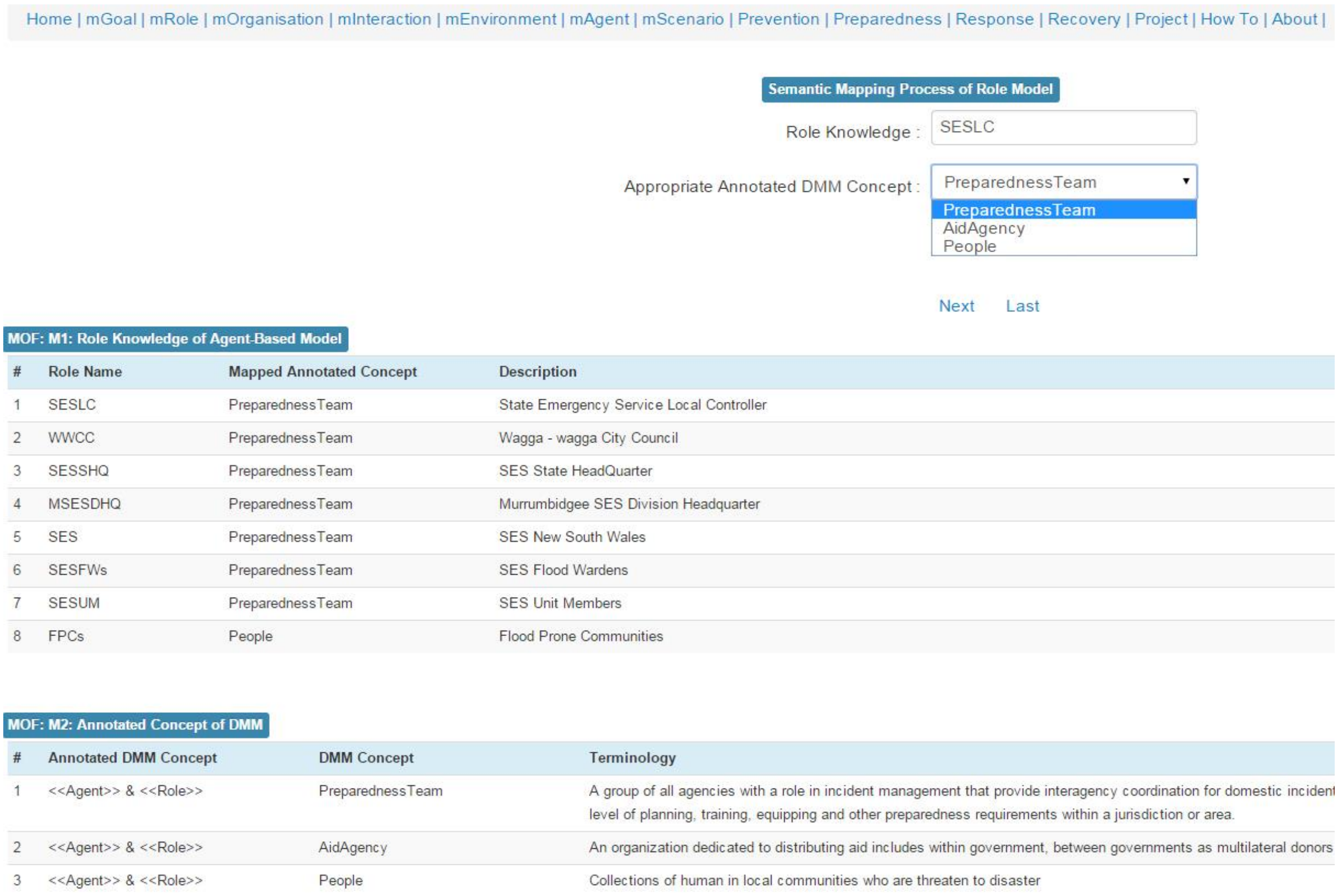

MOF: M1: Role Knowledge of Agent-Based Model

| # | Role Name | Mapped Annotated Concept | Description |
|---|---|---|---|
| 1 | SESLC | PreparednessTeam | State Emergency Service Local Controller |
| 2 | WWCC | PreparednessTeam | Wagga - wagga City Council |
| 3 | SESSHQ | PreparednessTeam | SES State HeadQuarter |
| 4 | MSESDHQ | PreparednessTeam | Murrumbidgee SES Division Headquarter |
| 5 | SES | PreparednessTeam | SES New South Wales |
| 6 | SESFWs | PreparednessTeam | SES Flood Wardens |
| 7 | SESUM | PreparednessTeam | SES Unit Members |
| 8 | FPCs | People | Flood Prone Communities |

MOF: M2: Annotated Concept of DMM

| # | Annotated DMM Concept | DMM Concept | Terminology |
|---|---|---|---|
| 1 | <<Agent>> & <<Role>> | PreparednessTeam | A group of all agencies with a role in incident management that provide interagency coordination for domestic incident level of planning, training, equipping and other preparedness requirements within a jurisdiction or area. |
| 2 | <<Agent>> & <<Role>> | AidAgency | An organization dedicated to distributing aid includes within government, between governments as multilateral donors |
| 3 | <<Agent>> & <<Role>> | People | Collections of human in local communities who are threaten to disaster |

*Figure 5. Mapping process of a role model and the annotated DMM concepts in Preparedness phase, an example*

Each layer consists of generic DM constructs which can be used to indexed DM knowledge during knowledge retrieval according to the disaster event context. These constructs can be mixed and matched between various DMPs according to each of the phases of Prevention, Preparedness, Response and Recovery. The conversion to this sophisticated representation is underpinned by an agent oriented analysis process. In this process, seven agent-oriented model templates are chosen and argued for in terms of the extant constructs within the DM knowledge in the existing DMPs and in terms of engendered characteristics of DM generally.

The DM metamodel constructs are applied to an intermediate set of agent based models that represent the DMP. This is supported by a semi-automated step to guide a DM practitioner on how to use apply the metamodel constructs. A tool is developed and described in the paper. A DM practitioner from SES is actually involved in overseeing the whole conversion process from DMPs to AO models to the metamodel-based repository. The paper successfully illustrates this approach by converting the actual flood DMP used by NSW SES at Wagga Wagga on the Murrumbidgee River in NSW to the metamodel based representation. The process focuses on resolving a set of key issues on how knowledge is typically stored, shared and accessed by participants and by communities in DM.

The work in this paper illustrated the effectiveness of the mapping process. The work successfully mapped an actual DMP into the metamodel representation. Future work will further develop the knowledge repository by converting all flood DMPs in NSW in collaboration with NSW SES. This will then be followed by an evaluation of the effectiveness of the repository for SES stakeholders and practitioners perspective.

# COPYRIGHT